\documentstyle[twocolumn,prb,aps]{revtex}
\begin{document} 
\title{
Superconducting Order Parameter in Bi-Layer Cuprates:
Occurrence of $\pi $ Phase Shifts in Corner Junctions
}
\author{D. Z. Liu, K. Levin, and J. Maly}
\address{
James Franck Institute, University of Chicago, Chicago, IL 60637
}

\address{\rm (Submitted to Physical Review B on 19 December 1994)}
\address{\mbox{ }}
\address{\parbox{14cm}{\rm \mbox{ }\mbox{ }
We study the order parameter symmetry in  bi-layer cuprates such as YBaCuO,
where interesting $\pi$
phase shifts have been observed in Josephson junctions. Taking models which
represent the measured spin fluctuation spectra of this cuprate, as well as
 more general models of Coulomb correlation effects, we classify the allowed
symmetries and determine their associated physical properties.
$\pi $ phase shifts are
shown to be a general consequence of repulsive interactions,
independent of whether a magnetic mechanism is operative. While it is known to
occur in d-states, this behavior can also be associated with (orthorhombic)
s-symmetry when the two sub-band gaps have opposite phase.
Implications for the
magnitude of $T_c$ are discussed.
}}
\address{\mbox{ }}
\address{\parbox{14cm}{\rm PACS numbers: 74.72. -h , 74.20.Mn, 74.50.+r, 74.62.
-c}}
\maketitle

\makeatletter
\global\@specialpagefalse
\def\@oddhead{REV\TeX{} 3.0\hfill Levin Group Preprint, 1994}
\let\@evenhead\@oddhead
\makeatother


The observation  in YBCO of unusual Josephson junction behavior
\cite{van-halingen,kirtley,mathai,ott} is  one of the
most important experimental results to emerge from the cuprate  literature in
recent years. Here in a  Josephson SQUID experiment the two junctions are
configured so that their normals lie along the a and b axes of the CuO$_2$
plane. This measurement has been  widely interpreted as support for a
d-symmetry of the order parameter, as well as for a magnetic mechanism for
superconductivity. In this paper we show that both of these inferences may be
inappropriate. For notational precision, throughout this paper
we use the terms s-  ( or d-  ) symmetry to correspond to states which have the
same  (or opposite) sign under a $\pi/2$ rotation. Thus the d-states under
consideration are not necessarily of the specific $d_{x^2-y^2}$ form.

The gap equation for bi-layer systems has been studied earlier in the context
of  a magnetic mechanism for superconductivity\cite{mazin,scalapino}.
 There it was observed
that the d-symmetric state of the single layer problem, is transformed to a
pair of in-phase d states on each of the two sub-bands, and that these compete
with a pair of out-of-phase s-states.  Here we take the problem to a greater
level of generality, establishing that this situation persists for a wide
class of repulsive interactions, which are  unrelated to the antiferromagnetic
spin fluctuation picture.
Alternate classes of the order parameter symmetry are also generated. These
correspond to in-phase s-states and out-of-phase d-states. We establish how
the relative stability of the two competing states is affected by changes in
bandstructure, orthorhombicity, and hole filling.

It should not be surprising that d-states have a more general origin beyond
the antiferromagnetic spin exchange models.  In a one layer cuprate, the
lattice symmetry requires that all gap states are either even (s-) or odd (d-)
under a $\pi/2$ rotation.
In bi-layer systems, these one layer states generalize naturally to a pair of
even or odd, in-phase or out-of-phase  states, associated with each of the two
sub-bands.  Thus, as one of two allowed states,
d-symmetry should  be widespread, and independent of the microscopic details
of the model.

In the presence of both intra- and inter-layer interactions ($V_{\parallel}$
and $V_{\bot}$), the weak coupling BCS gap equation  becomes a set of
coupled equations for the  gaps on each of the sub-bands.  It is simpler to
write the gap equations in terms of the two sub-band gaps $\Delta_+, \Delta_-$
rather than the
intra-layer  ($\Delta_{\parallel}$)
and inter-layer  ($\Delta_{\bot}$) components. These are related via
the unitary transformation
which diagonalizes the Hamiltonian.
In this model $t_{\bot}$ is the matrix element for hopping between layers; $t$
and $t'$
refer to  the first and second nearest neighbor in-plane hopping which may
contain orthorhombic effects.
On site Coulomb effects $U$ are assumed to enter via a renormalization of the
bandstructure parameters as  shown by Si {\it et al} \cite{si-qm}.

Following the usual procedure\cite{intergap}, the gap
equations become
\begin{mathletters}
\begin{eqnarray}
\Delta_++\Delta_- &= & -\sum_{{\bf
q}'}\frac{V_{\parallel}\Delta_+}{2E_+}\tanh\left(\frac{E_+}{2T}\right)
\nonumber \\
 & & \mbox{ }
-\sum_{{\bf
q}'}\frac{V_{\parallel}\Delta_-}{2E_-}\tanh\left(\frac{E_-}{2T}\right) \\
\Delta_+-\Delta_- &= & -\sum_{{\bf
q}'}\frac{V_{\perp}\Delta_+}{2E_+}\tanh\left(\frac{E_+}{2T}\right)
\nonumber \\
 & & \mbox{ }
+\sum_{{\bf
q}'}\frac{V_{\perp}\Delta_-}{2E_-}\tanh\left(\frac{E_-}{2T}\right)
\end{eqnarray}
\end{mathletters}
where the superconducting quasi-particle energies are
$E_{\pm}=\sqrt{\epsilon_{\pm}^2+\Delta_{\pm}^2}$, where
\begin{eqnarray}
\epsilon_{\pm}& =&-2t[\cos(q_xa)+\cos(q_yb)] \nonumber \\
 & & \mbox{ }+4t'\cos(q_xa)\cos(q_yb)\pm t_{\perp}
\end{eqnarray}
It follows that each of the two sub-band gaps can be written in terms of the
parallel and perpendicular components as
\begin{equation}
 \Delta_{\parallel}= (\Delta_+ + \Delta_-)/2,
\mbox{ }\Delta_{\perp}= (\Delta_+ -\Delta_-)/2 \label{bily-q4}
\end{equation}

In the case of a magnetic pairing mechanism, the two interactions are related
to  components of the dynamical spin susceptibility.  This susceptibility has
been  calculated for the bi-layer cuprate YBCO\cite{si-qm,zha},
in the strong  $U$
limit.  For realistic Fermi surface shapes, and moderate in-plane and out of
plane exchange interactions,
the results are in reasonable agreement with neutron
experiments\cite{tranquada}.
Because the in-plane magnetism is not independent of
inter-plane effects any proper treatment of spin fluctuation induced
superconductivity should incorporate  both components. A reasonable
approximation of both the theoretical results\cite{zha} and the experimental
neutron data\cite{tranquada}  is to take
\begin{mathletters}
\begin{eqnarray}
V_{\parallel} &= &g_{\parallel}^2 \chi_{\parallel}({\bf q} - {\bf q}')
\label{bily-q5a} \\
V_{\perp} & =& g_{\perp}^2 \chi_{\perp}({\bf q} - {\bf q}') \nonumber \\
 &=& -g_{\perp}^2 \chi_{\parallel}({\bf q} - {\bf q}') \label{bily-q5b}
\end{eqnarray}
with
\begin{equation}
    \chi_{\parallel}(k_x ,  k_y ) = \frac{1}{[1+J_o(\cos k_xa+\cos k_yb)]^2}
\label{bily-q5c}
\end{equation}
\end{mathletters}
and $J_o\approx 0.3$.
Here we have absorbed overall
coefficients into the coupling constant prefactors $g_{\parallel}$ and
$g_{\perp}$.  This model is similar to that used in Ref.5, except that
we have assumed an arbitrary relation between the magnitudes of two
superconducting coupling constants, which are taken to be the same in Ref.5.
Moreover, we fit the dynamical susceptibility to neutron, rather
than  NMR data\cite{Eugene}.

It may be noted from Eq.(\ref{bily-q5c}) (as well as
experimental neutron data\cite{tranquada})
 that the antiferromagnetic fluctuations show up as a weak
peak around $(\pi/ a, \pi/a)$ with short coherence lengths, suggesting that
the system is far from any real instability; thus high $T_c$ is difficult to
explain\cite{Eugene}.
Consequently we explore more general mechanisms by extending
Eqs.
(\ref{bily-q5a}) and (\ref{bily-q5b}) to the case where the overall signs are
unconstrained and the ${\bf q}-{\bf q}'$ peaks occur at arbitrary
wave-vector with arbitrary peak width
$J_o$.
We define
\begin{eqnarray}
   V_{\parallel, \perp} = \frac{\lambda_A,\lambda_B}{[1-J_o(\cos
(k_x\pm Q_x)+\cos
(k_y\pm Q_y))]^2}
\end{eqnarray}
where all signs are summed over. We
divide our numerical analysis into four distinct cases in which  either
the  inter-layer or the intra-layer interaction dominates and in which the
respective  interaction is repulsive ($\lambda > 0$)  or attractive ($\lambda<
0$).The model is viewed as a general representation of
pairing mechanisms of the electronic and phononic variety. Since the
former usually derives from a generalized susceptiblity, and the latter
from a phonon propagator,there are no sign changes as a function of
momentum transfer. This poses important constraints on the allowed
superconducting states. Our results\cite{phases}
are summarized in Figures 1 and 2 for the  case of
characteristic wave-vectors ${\bf q}-{\bf q}'$ along $(\pi / a , \pi /a)$ and
along $(\pi /a , \pi/2a)$.
The latter wave-vector illustrates the behavior away from the antiferromagnetic
instability model, in order to show the generality of our results.
Plotted in these figures are the form of the gap functions in the two regimes.
The figures on the left (right) in each box correspond to intra-layer
(inter-layer) dominated behavior. Our
conclusions from both Figures 1 and 2 may be succinctly summarized.
We find that d-symmetry is
associated with repulsive and s-symmetry with attractive interactions. In
phase gap behavior occurs when the intra-layer interaction is the
larger; out of  phase behavior arises in the opposite case. This phase
dependence  can be deduced from Eqs.(\ref{bily-q4}). In the case of dominant
inter-layer effects, $|\Delta_{\bot}| > | \Delta_{\parallel}|$.  This will
occur when $\Delta_+$ and $\Delta_-$ have opposite signs.

To establish the generality of
these results,  we have varied the Fermi surface shape ( via the ratio of $t'
/ t$ ), the position of the Fermi  energy or  hole filling and the width $J_o$
of the peak structure. These variations introduce only
quantitative but not qualitative changes in the physical picture shown in the
two figures.

It is important to note from  Figs.\ref{bily-f1}(a)-(c) and
\ref{bily-f2}(a)-(c)   that the out of phase s- and  d-wave
states will exhibit $\pi$ phase shifts in a corner SQUID experiment
\cite{van-halingen,kirtley,mathai,ott}. This
corresponds to a change in sign of the "sum " order parameter
$\Delta_{\parallel}$ of Eqs.(\ref{bily-q4})  upon varying
from 0 to $\pi/2$.   While not a general feature of all solutions,  its
presence requires  (a modest amount of ) orthorhombicity.   The observation of
$\pi$ phase shifts in  bi-layer cuprates is thus not as strong a
constraint on the order parameter symmetry as in  one layer materials.  All
orthorhombic states which exhibit these $\pi$ phase shifts will  also show
finite c-axis tunneling in untwinned crystals\cite{dynes}.
 However, twinning effects (if
they  average  fully over the a and b-axes), will lead to a cancelation of
Josephson coupling,  whenever the corner SQUID experiment  has the observed
$\pi$ phase shift.

It should be stressed that the  out of phase s-states have the additional
advantage as a  candidate gap symmetry, over d-states (in or out of phase), of
being relatively  insensitive to impurity effects. In addition,
this  state can be compatible with neutron
experiments\cite{tranquada},
 which show no nodal signature. In contrast to experiment, because
nodes are not present, at least in the clean limit, power laws  in
thermodynamical properties\cite{bonn} are not expected.

We have searched for nodal behavior in these s-wave states with some care,
since  there is recent photoemission evidence\cite{argonne}
 to suggest that they may exist
in bi-layer BISCO.   Several observations are important to note in this
context. (1)  As the inter-layer hopping $t_{\bot}$ becomes small the
magnitudes of the two  gaps become equal and they are less  able to
respond to orthorhomicity by producing gap anisotropy.  Consequently BISCO
2212, which is believed to have a very small $t_{\bot}$, would be unlikely  to
exhibit  nodal  s-wave behavior.  (2) The eight node s-state which has been
conjectured as a  candidate for BISCO\cite{argonne}
 appears quite generally as a meta-stable
state whose solution has a lower  $T_c$ than the nodeless s-wave or  (four
node) d-wave symmetric gap.  Within the manifold of meta-stable states, the
more nodes, the

\onecolumn

\begin{figure}
 \vbox to 5cm {\vss\hbox to 6.5cm
 {\hss\
   {\includegraphics{/home/ldz/tek/paper/psd04/f1a.ps}
   }
   {\includegraphics{/home/ldz/tek/paper/psd04/f1b.ps}
   }
  \hss}
 }
 \vbox to 5.0cm {\vss\hbox to 6.5cm
 {\hss\
   {\includegraphics{/home/ldz/tek/paper/psd04/f1c.ps}
   }
   {\includegraphics{/home/ldz/tek/paper/psd04/f1d.ps}
   }
  \hss}
 }
\caption{
Superconducting gap for interactions peaked at ${\bf q}-{\bf q}'=(\pi, \pi )$,
for the case of attractive and repulsive intra-layer ( $\lambda_A$) and
inter-layer ($\lambda_B$) interactions. Figures on left (right) are for intra-
(inter-)
layer dominated regimes.
\label{bily-f1}}
\end{figure}

\begin{figure}
 \vbox to 5.0cm {\vss\hbox to 6.5cm
 {\hss\
   {\includegraphics{/home/ldz/tek/paper/psd04/f4a.ps}
   }
   {\includegraphics{/home/ldz/tek/paper/psd04/f4b.ps}
   }
  \hss}
 }
 \vbox to 5.0cm {\vss\hbox to 6.5cm
 {\hss\
   {\includegraphics{/home/ldz/tek/paper/psd04/f4c.ps}
   }
   {\includegraphics{/home/ldz/tek/paper/psd04/f4d.ps}
   }
  \hss}
 }
\caption{
Superconducting gap for interactions peaked at ${\bf q}-{\bf q}'=(\pi, \pi/2)$.
\label{bily-f2}}
\end{figure}

\twocolumn
\noindent
lower  is the $T_c$.
(3) We have studied solutions to a separable pairing potential model
in  which the susceptibility of Eq.(\ref{bily-q5c}) is replaced by a product
of cosine terms:  $\cos (q_xa) \cos (q_x' a) + \cos (q_y b) \cos (q_y' b)$,
since  it was
 speculated\cite{fedro} that this potential would give rise to
an 8 node s-state.   We find that
d-wave states  arise naturally in this model as well, and they
are always
more stable than s-wave states.

\begin{figure}
 \vbox to 5.0cm {\vss\hbox to 6.5cm
 {\hss\
   {\includegraphics{/home/ldz/tek/paper/psd04/f3.ps}
   }
  \hss}
 }
\caption{
Effect of inter-layer correlation on the superconducting transition
temperature $T_c$ . ($\bigtriangleup$) indicate the  $(d, d)$ states
and ($\bigcirc $)
represent the $(s, -s)$ states. We consider different shapes of Fermi surface
 and Fermi energies -- orthorhombic lattice with next-nearest-neighbor hopping,
$E_F=-0.2$ (dotted line),
$E_F=-0.1$
(solid line); and tetragonal lattice with $t'=0$,
$E_F=-0.6$ (dot-dashed line),
$E_F=-0.4$ (dashed line).
\label{bily-f3}}
\end{figure}

In Fig.\ref{bily-f3} is plotted the dependence of the transition temperature
on the inter-plane coupling constant $T_c$ for the magnetic pairing model of
Fig.\ref{bily-f1}(a),  for two different Fermi surface
shapes\cite{si-qm}
 corresponding to YBaCuO (with two hole concentrations as well )  and
LaSrCuO ($t' = 0$ ). Here, because we use a weak coupling approximation,  the
absolute  values of $T_c$ are not meaningful.  However, the relative changes
with different  parameterizations are expected to be accurately captured.
The circles correspond to the out-of-phase s-states and triangles to in-phase
d-states. As expected,  for  sufficiently small inter-plane coupling the
d-wave state is the more stable;  however depending on the bandstructure and
Fermi surface shape, this state may be readily de-stabilized to a pair of s
states by the  introduction of a very small  inter-plane interaction. This
reflects the general result that (nodeless) s-wave states are able to take
better advantage of the superconducting interaction than can d-wave states,
which require a cancellation of positive and negative terms to satisfy  the
gap equation.  Note that in the LaSrCuO model
the $T_c$'s are generally higher as a
consequence the better Fermi surface nesting along the direction of the
wave-vector $(\pi , \pi)$. For each parameter set, the
various curves tend to coalesce at higher
$\lambda_B$, where the (s-like) states are found to be more isotropic.  This
is a consequence of the fact  that these isotropic states are not able to
utilize the Van Hove singularity effect, which  is relatively more important
for d$_{x^2-y^2}$ state.

In summary, by solving the gap equation for bi-layer models with general
repulsive interactions,  we find that d$_{x^2-y^2}$ states arise quite
generally
and are not  uniquely associated with wave-vector structure along the
antiferromagnetic direction. Moreover, we have established that $\pi$ phase
shift
behavior, which is often cited as the strongest
evidence for d-wave pairing  can also be associated with (orthorhombic)
s-symmetry when the two sub-band gaps have opposite phase. This state has
some advantages over d-states in large part because of the relatively small
sensitivity  of $T_c$ to non-magnetic impurities. An important conclusion from
our analysis is that there  are always competing states in bi-layer systems,
and that the order parameter symmetry would be expected to vary from  cuprate
to cuprate as well as  within  a given  cuprate  class at different hole
concentrations. One can conclude that the Josephson junction data,  in
particular,  provide strong  evidence for superconductivity mediated by some
form of repulsive interaction. On the other hand, these collected observations
(in bi-layer cuprates) weaken the often cited  support  for theories of
spin fluctuation mediated  superconductivity.

We acknowledge useful conversations with A. Leggett, M. Norman
and P. Wiegmann, and
correspondence with V. Emery.  This
work was supported by the NSF through the Science and Technology Center
for Superconductivity (DMR 91-20000).


 Note Added:

After this manuscript was submitted we received a preprint from K. Kuboki and
P. A. Lee, in which an RVB description of bi-layer superconductivity
was used to
infer a spontaneous breaking of tetragonal symmetry. This $s,d$ mixing
will not occur
in the present model, as a result of the free energy form which contains
a quadratic, rather than quartic, mixing of the in- and out- of- plane gaps.

\begin{references}
\bibitem{intergap} We do not include any effects of inter-band pairing,
which we find are unphysical because they introduce additional transition
temperatures.
\bibitem{phases} Besides those shown in the figures, other s- and d- phases
may appear for restricted regions of our parameter space. As a result of
the Van Hove effect the $d_{x^2-y^2}$ form of the d- representation tends to be
more robust.
\end{references}
\end{document}